\documentclass[%
 reprint,
nofootinbib,
nobibnotes,
amsmath,amssymb,
aps,
prd,
]{revtex4-1}

\usepackage{graphicx}
\usepackage{dcolumn}
\usepackage{bm}
\usepackage{url	}

\usepackage{hyperref}
\usepackage{cleveref}
\usepackage[dvipsnames]{xcolor}

\hypersetup{
    colorlinks,
    citecolor=[rgb]{0.16,0.67,0.53},
}

\newcommand*\apjl{Astrophys.~J.~Lett}

\newcommand*\apss{Ap\&SS}

\newcommand*\jcap{J. Cosmology Astropart. Phys.}

\newcommand*\mnras{Mon.~Not.~Roy.~Astron.~Soc.}

\newcommand*\pasp{PASP}

\begin{document}


\title{Constraints on field flows of quintessence dark energy}

\author{Federico Tosone}
 \email{federico.tosone@roma2.infn.it}
\author{Balakrishna S. Haridasu}
 \email{haridasu@roma2.infn.it}
\author{Vladimir V. Lukovi\'c}
 \email{vladimir.lukovic@roma2.infn.it}
\author{Nicola Vittorio}
 \email{nicola.vittorio@roma2.infn.it}
\affiliation{Dipartimento di Fisica, Universit\'a di Roma "Tor Vergata", Via della Ricerca Scientifica 1, I-00133, Roma, Italy
}%
\altaffiliation{Sezione INFN, Universit\'a di Roma "Tor Vergata", Via della Ricerca Scientifica 1, I-00133, Roma, Italy}

\begin{abstract}
The quest for understanding the late-time acceleration is haunted by an immense freedom in the analysis of dynamical models for dark energy in extended parameter spaces. Often-times having no prior knowledge at our disposal, arbitrary choices are implemented to reduce the degeneracies between parameters. We also encounter this issue in the case of quintessence fields, where a scalar degree of freedom drives the late-time acceleration. In this study, we implement a more physical prescription, the \textit{flow} condition, to fine-tune the quintessence evolution for several field potentials. We find that this prescription agrees well with the most recent catalogue of data, namely supernovae type Ia, baryon acoustic oscillations, cosmic clocks and distance to last scattering surface, and it enables us to infer the initial conditions for the field, both potential and cosmological parameters. At $2\sigma$ we find stricter bounds on the potential parameters $f/m_{pl}>0.26$ and $n<0.15$ for the PNGB and IPL potentials, respectively, while constraints on cosmological parameters remain extremely consistent across all assumed potentials. By implementing information criteria to assess their ability to fit the data, we do not find any evidence against thawing models, which in fact are statistically equivalent to $\Lambda$CDM, and the  freezing ones are moderately disfavoured. Through our analysis we place upper bounds on the slope of quintessence potentials, consequently revealing a strong tension with the recently proposed swampland criterion, finding the 2$\sigma$ upper bound of $\lambda \sim 0.31$ for the exponential potential.
\end{abstract}

\maketitle

\section{Introduction}
The successful detection of the late-time accelerated expansion of the Universe \citep{Riess98,Schmidt98} made the nature of dark energy one of the most engaging problem in modern cosmology. The identification of the energy density of a cosmological constant as the missing component became a paradigm of the standard model. Nonetheless the value inferred from cosmic acceleration cannot lay the groundwork on quantum field theory, being too small in comparison to the prediction made from particle physics \citep{Carroll00,Martin12}. One popular alternative to the cosmological constant is that dark energy corresponds to the expectation value of the energy density of a scalar field evolving under a potential \citep{Ratra87, Wetterich88}. The observed energy density is thus a dynamical vacuum that eventually will approach a zero minimum, and it is the fifth component of the cosmos in addition to cold dark matter, baryons, photons and neutrinos, thus it was dubbed quintessence \citep{Caldwell97}.\\ More exotic scalar fields were proposed as driving components of the late time speed-up of the Universe, such as k-essence \citep{Chiba00,Armendariz01}, phantom energy \citep{Caldwell99}, tachyons \citep{Padmanabhan02}, ghost condensates \citep{ArkaniHamed03,Piazza04}. For comprehensive reviews about dark energy theories see \cite{Copeland06,Bamba12,Ishak18,Heisenberg18}.

Another possible solution to the problem is that gravity weakens on cosmological scales, resulting in cosmic acceleration (see \cite{Joyce16} for a review on the distinction from dark energy). Indeed modified gravity (MG) theories, though might be indistinguishable from dark energy at the level of cosmic expansion alone, they leave distinct features on the large-scale structure \citep{Linder05}, which can be detected by various observational probes \citep{Koyama15}. Given the current observational status, the parameter space related to these models has been probed by the means of phenomenological approaches, with the standard model being favoured by data, though there are indications for a dynamical evolution \citep{Zhao17_nature, Sahni14}. Recently the broad scenario of MG was severely narrowed down by the detection of the electromagnetic counterpart of a gravitational wave signal, because of the stringiest bounds on the speed of the graviton \citep{Lombriser16,Sakstein17,Ezquiaga17}. In addition the clustering of galaxies on non-linear scales shows already strong evidence against deviations from general relativity \citep{He18}. The precise forthcoming data from surveys like DESI and EUCLID \citep{Levi13,Amendola16} could be decisive in unravelling the nature of late-time acceleration. In this respect quintessence models are favoured by recent observations, and several studies place constraints on it by the means of numerical approaches, approximate solutions or parametrizations of the evolution of the field \citep{Barreto15, Lonappan17, Zhai17, Durrive18, Sangwan18, Yang18}. Because of various assumptions made in the analysis, such as those on the initial conditions of the field or on the parameters of the potential, we regard these results as not the most general. However these assumptions are often needed to fine tune quintessence evolution \citep{Carroll00} or to reduce the parameter space of the  models and ease its exploration by the means of a Monte Carlo Markov Chain (MCMC) analysis. In the current study we improve on previous works by implementing a statistical framework to constrain several quintessence models. While the rich dynamics of quintessence field depends mainly on the potential (see \cite{Tsujikawa13} for an introduction to quintessence dynamics), we are specifically concerned with the analysis of these models in an unbiased way, without resorting to arbitrary priors on the parameters of the potentials nor on initial conditions. We analyse the potentials which have a well defined evolution in the configuration space of the field \citep{Caldwell05,Marsh14}, and which are well known in the literature as plausible quintessence candidates motivated by high energy physics. For these classes we implement a formalism based on fine-tuning a non linear combination of initial conditions and potential parameters in the matter era, according to the \textit{flow} condition \citep{Cahn08}. This prescription was shown to allow a natural dynamical evolution of the field, and here we develop the Bayesian formalism to exploit it in a MCMCs analysis. We have tested this framework against the most updated catalogue of observables, including Baryon Acoustic Oscillations (BAO) \citep{Eisenstein05}, distance to last scattering from Cosmic Microwave Background measurements (CMB) \citep{Planck15_DE}, Supernovae Type Ia (SN) \citep{Betoule14} observations and Cosmic Clocks (CC) \citep{Jimenez01}.\\
We also complement the parameter inference from our analysis with model selection utilising the information criteria. In this regard, we try to evaluate the statistical evidence for or against some of the high energy physics quintessence potentials we considered. As our analysis readily provides constraints on the quintessence fields, we also comment on the recently proposed swampland criteria in the context of string theory \citep{Obied18}. Current observations are able to probe swampland parameter space \citep{Agrawal18}, and provide limits on the available region \citep{Heisenberg18,Akrami18}. In this work we extend the discussion based on our improved constraints on the same parameter space.\\
The paper is structured as follows: in \cref{sec:Theory1} we summarize the dynamics of quintessence models for different choices of potential, and in \cref{sec:Theory2} we discuss the associated \textit{flow} dynamical condition. In \cref{sec:Data} we list the observations utilised, whose likelihoods are defined in \cref{sec:Methods} together with the implementation of the \textit{flow} condition as a prior. Our results are outlined in \cref{sec:Results}, and we conclude in \cref{sec:Conclusions}.

\section{Theory}
\subsection{Quintessence fields}
\label{sec:Theory1}
The standard action with the inclusion of a quintessence field can be written as,
\begin{equation}\label{eq:Action}
\mathcal{S} = \int \sqrt{-g} d^4 x \{ \frac{m^2_{pl}}{2} R  -\frac{ \partial^{\mu} \phi \partial_{\mu} \phi}{2} - V(\phi)  \} + \mathcal{S}_{m},
\end{equation}
where $m^2_{pl} = (8 \pi G)^{-1} $ is the reduced Planck-mass squared, $V(\phi)$ is the potential of the scalar field, and $\mathcal{S}_m$ contains the contribution of matter fields. In order to study the background expansion of the Universe we consider the average value of the field $\phi$, which is a function of time alone under the standard assumptions of isotropy and homogeneity, and the corresponding energy-momentum tensor is diagonal $T^{\nu}_{\mu} = diag(-\rho_{\phi}, p_{\phi} , p_{\phi}, p_{\phi})$. Adopting the perfect fluid prescription, the equation of state (EoS) for the field is given by,
\begin{equation}\label{eq:EoS}
w_{\phi} = \frac{p_{\phi}}{\rho_{\phi}} = \frac{\dot{\phi}^2-2 V(\phi)}{\dot{\phi}^2+2 V(\phi)}.
\end{equation} 
\indent The EoS dictates the dynamical evolution of the field, and in-turn contributes to the Hubble flow through
\begin{equation}\label{eq:Hubble}
\frac{H(a)^2}{H_0^2} \equiv  E(a)^2 = \left[ \frac{\Omega_{r}}{a^4} + \frac{\Omega_{m}}{a^3} + \frac{\Omega_{k}}{a^2} + \Omega_{\phi} f(a)  \right], 
\end{equation}
where $H(a)$ is the Hubble rate written in terms of the scale factor $a$, $\Omega_x = \rho_{x0}/\rho_{crit0}$ is the current energy density of the $x$ fluid component normalized with respect to the current critical density of the Universe $\rho_{crit0} = 3 H^2_0 m^2_{pl}$, and subscript `0' corresponds to the quantities measured today ($a = 1$). We use the convention that $\Omega_{\phi} \equiv \Omega_{\phi0}$ when the time dependence is not explicitly mentioned, and $w_0 \equiv w_{\phi}(a=1)$. $f(a)$ is obtained from continuity equation of the field as
\begin{equation}\label{eq:f(a)}
f(a) = \exp{\left[ 3 \int_{a}^{1} \frac{d \tilde{a}}{\tilde{a}} (1+w_{\phi}(\tilde{a}))\right]}.
\end{equation}
\indent In this work we investigate only the flat ($\Omega_k = 0$) scenario for which the transverse comoving distance reads as
\begin{equation}\label{eq:D_M}
D_{M} = \frac{c}{H_0} \int_0^{z} \frac{dz}{E(z)}.
\end{equation}
\indent Finally, the evolution of the quintessence field is obtained by solving the initial value problem for the non-linear system including \cref{eq:Hubble} in concert with Klein-Gordon equation
\begin{equation}\label{eq:KleinGordon}
\ddot{\phi} + 3 H \dot{\phi} + V_{,\phi}(\phi) = 0,
\end{equation}
requiring the specification of initial conditions \{$\phi_{i}$, $\dot{\phi}_{i}$\}, the functional form of $V(\phi)$ and the background components. The Klein-Gordon equation implies that the field is rolling down the potential with an acceleration directly proportional to the steepness  $(V_{,\phi})$, against a friction term driven by the background expansion $(3 H \dot{\phi})$. Varied choices for the potential can give rise to rich spectrum of dynamical behaviour, which we briefly outline here.\\
Among quintessence candidates for dark energy, the exponential potential $V(\phi) = V_0 e^{-\lambda \phi/m_{pl}}$ (EXP) was one of the first to be considered, as it arises naturally in higher dimensional theories \citep{Ferreira97}. More recently it has been proposed as a scale-invariant extension of the Standard Model non-minimally coupled to unimodular gravity \citep{Rubio18}. EXP potential exhibits the attractive property of a cosmologically scaling solution when $\lambda^2 >2$ \citep{Ferreira97}, wherein dark energy density scales with the dominant background component according to \citep{Tsujikawa13},
\begin{equation}
\label{eq:scaling1}
\Omega_{\phi}(a) \sim \frac{3(1+w_{bg})}{\lambda^2}
\end{equation}
where $\Omega_{\phi}(a)$ is the scalar field energy density at time $a$, normalized with respect to the critical density $3 H^2(a) m^2_{pl}$, and $w_{bg}$ is the EoS of the dominant background component, which could be either radiation ($w_r = 1/ 3$) or matter ($w_m = 0$). \Cref{eq:scaling1} was considered as a possible way to ameliorate the coincidence problem \citep{Carroll00}, but it was shown that there is not a transition to a de Sitter solution at late-time, namely a solution where dark energy density is dominant and the Universe accelerates, unless one considers $\lambda < \sqrt{2}$ \citep{Copeland97}. In this case the field loses its scaling properties, and it follows a thawing evolution \citep{Caldwell05}. The thawing behaviour is common to all quintessence fields that are slowly rolling in a sufficiently flat potential: while they remain frozen with EoS $w_{\phi} \simeq -1$ during the radiation and matter dominated eras, they start to roll towards the late-time as the Hubble friction reduces and eventually become the dominant species \citep{Scherrer07}. Thawing evolution requires fine-tuning in order to be consistent with observations, thus these models can closely mimic $\Lambda$CDM to the point of being indistinguishable \citep{Linder15}. Nonetheless, there are thawing fields which are well motivated in the context of high energy physics, such as the Pseudo Nambu-Goldstone Boson (PNGB). PNGBs are ultra-light axions with a mass $m_{\phi} \sim H_0$, that arise in the context of explicit symmetry breaking. They have the attractive property of their small mass being protected against radiative corrections, and also couplings to Standard Model fields are suppressed \citep{Carroll98}, which makes them physically motivated quintessence candidates. One expects that their low-energy effective field theory potential takes the form \citep{Frieman95},
\begin{equation}\label{eq:PNGB}
V_0  \left[\cos{\left(\frac{\phi}{f}\right)} +1 \right],
\end{equation}
where $V_0$ is the normalization of the potential and $f$ is the shift symmetry scale. It is worth mentioning that potentials of this form naturally arise in the context of string theory \cite{Kamionkowski14}, and can also be motivated in a supergravity scenario \citep{Chiang18}. In our study we also consider a power law potential (PL) $V(\phi) \sim \phi^{n}$ as another viable quintessence potential. Power law potentials are ubiquitous in field theories, and their phenomenological evolution is also of thawing kind \cite{Linder15}.\\ 
As opposed to thawers, a potential which could drive the late-time acceleration is an inverse power law (IPL), with $V(\phi)\sim \phi^{-n}$ \citep{Ratra87}. IPL is representative of a whole class of potentials with the EoS of the field tracking the background equation of state \citep{Steinhardt99} as,
\begin{equation}
\label{eq:scaling}
w_{\phi} \simeq \frac{w_{bg}-2(\Gamma-1)}{1+2(\Gamma-1)},
\end{equation}
and $\Gamma$ is defined as
\begin{equation}
\Gamma = V \frac{V_{, \phi \phi}}{ V^2_{, \phi} }.
\end{equation}
\indent In the case of an IPL potential $\Gamma = \frac{n+1}{n}$ is constant, such that $w_{\phi}$ is approximately a constant as well in the matter era. The Supergravity potential (SUGRA) is yet another tracker, which in addition is consistent with super-Plankian values of the field \citep{Brax99} and is given as,
\begin{equation}
V_0 \phi^{-n} e^{ \frac{ \phi^2 }{2 m_{pl}^2} }.
\end{equation}
\indent Tracking models are commonly referred as freezing quintessence, and the corresponding dynamical evolution is different from $\Lambda$CDM only at high redshift, in the tracking regime, while today the field is expected to slowly roll such that it approaches the de Sitter solution $w \sim -1$.  Originally they were proposed in order to solve the coincidence problem, as their dynamics admit a wide choice of initial conditions \{$\phi_{i}, \dot{\phi}_{i}$\} consistent with their tracking properties, yet they require fine-tuning of the energy density in order to be consistent with late-time observations. The differentiation between thawing and freezing models is well drawn, and it corresponds to distinctive priors in the configuration space of the field \citep{Caldwell05}, which could be a valuable means to disclose the nature of the field giving rise to the late-time acceleration.\\
A model which eludes this distinction is the double exponential potential (DEXP) \citep{Barreiro99},
\begin{equation}\label{eq:DEXP}
V_0 \left( e^{ -\lambda_1 \phi/m_{pl}} + \mu e^{ -\lambda_2 \phi/m_{pl}}  \right),
\end{equation}
though it can be considered to be of the freezing kind, DEXP is a scaling model as it exploits the scaling regime of EXP at early-times for $\lambda_1 \gg \sqrt{2}$ and $\lambda_1 \gg \lambda_2$, while it shows a de Sitter behaviour at late-times for $\lambda_2 < \sqrt{2}$. Here, $\mu$ is an additional free parameter in the analysis, which implies that the two potentials may be normalized differently. We list these potentials in \cref{tab:potentials}, alongside with their dynamical behaviour in the configuration space of the field. Notice that apart from SUGRA, all of them have a zero minimum for the potential.\\
\\
We rewrite the Klein-Gordon equation in a formalism which is independent of the explicit definition of the Hubble rate, so that it allows one to decouple the scalar field evolution from the background cosmology. This is accomplished by writing \cref{eq:KleinGordon} in terms of the independent variable $\ln{a}$, of the time-dependent normalized energy density $\Omega_{\phi}(a)=\rho_{\phi}(a)/\rho_{crit}(a)$ and of the EoS $w_{\phi}(a)$, in a Universe made of matter and dark energy fluids (the same formalism is adopted by \cite{Huterer06}),
\begin{equation}\label{eq:KleinGordon2}
\phi'' + \frac{3}{2} (1- w_{\phi}(a) \Omega_{\phi}(a)) \phi' + \frac{V_{,\phi}}{V} \Omega_{\phi}(a) (1-w_{\phi}(a)) = 0,
\end{equation}
where a prime denotes the derivative with respect to $\ln{(a)}$, and we used the units $m_{pl}=1$. The solution of \cref{eq:KleinGordon2} is still an initial value problem, namely to perform the integration we need to specify the initial conditions $\{ \phi_i, \phi'_i \}$, and the potential. For convenience we express the EoS and the normalized energy density in terms of the auxiliary functions $A$ and $B$
\begin{equation}\label{eq:wFun}
w_{\phi} = \frac{2B(1+A)}{2A+B} -1,
\end{equation}
\begin{equation}\label{eq:OmegaFun}
\Omega_{\phi} = \frac{2A+B}{2(1+A)},
\end{equation}
where we defined $A \equiv V(\phi)/\rho_m(a)$. In turn $A$ can be easily expressed in terms of the scalar field observables at the initial time of integration (index $i$), so that the definitions of $A$ and $B$ read as
\begin{equation}\label{eq:AB}
\begin{aligned}
A&=\frac{V(\phi)}{V(\phi_i)} \frac{\Omega_{\phi}(a_i)(1-w_{\phi}(a_i))}{2(1-\Omega_{\phi}(a_i))} \left(\frac{a}{a_i} \right)^3, \\
B &= \frac{1}{3}\left( \frac{d \phi}{d \ln{a} } \right)^2
\end{aligned}
\end{equation}
Notice that we do not need an explicit normalization constant $V_0$ to evolve the field, since \cref{eq:AB,eq:KleinGordon2} are independent of it.

{\renewcommand{\arraystretch}{1.25}%
	\begin{table}
		
		\centering
		\caption{
			The potentials considered in our analysis are summarized here. They are characterized by freezing (fr) or thawing (th) behaviour in configuration space}
		\label{tab:potentials}
		\begin{tabular}{lccc} 
			\hline
			Model & $V(\phi)/V_{0}$  & Reference \\
			\hline
			EXP (th) & $e^{-\lambda \phi}$ & \cite{Copeland97} \\
			PNGB (th) &  $\cos{\left(\frac{\phi}{f}\right)} +1$  & \cite{Frieman95} \\
			PL (th) & $\phi^n$  &  \cite{Linder15} \\
			IPL (fr) & $\phi^{-n}$ &  \cite{Ratra87} \\
			SUGRA (fr) &  $ \phi^{-n} e^{\frac{\phi^2}{2}}$ & \cite{Brax99} \\
			DEXP (fr) & $e^{ -\lambda_1  \phi} + \mu e^{ -\lambda_2  \phi} $ & \cite{Barreiro99} \\
			\hline
		\end{tabular}
	\end{table}
}

\subsection{Flow prior}
\label{sec:Theory2}
The nature of a quintessence field is mostly determined by the assumed potential, however the dynamical scenario is still flexible enough to cover most of the configuration space of the field \citep{Linder06}, and it is affected by parameters of the potential and initial conditions. Apparently there is no guidance to set priors on the parameter space, except for requiring a model to be phenomenologically consistent with observations. Admittedly, there is a conserved dynamical quantity for both freezing and thawing models, and it can be inferred from first principles. A thawer is frozen by Hubble friction in the long matter era, so that one can solve the Klein-Gordon equation in a perturbative way to second order in this limit. This results in the following precise prescription on a non-linear combination of physical observables \citep{Cahn08}
\begin{equation}\label{eq:Flow}
F(a) =  \frac{1+w_{\phi}(a)}{\lambda^2(\phi) \Omega_{\phi}(a)} = \frac{4}{27}.
\end{equation}
\indent Where $\lambda = V_{,\phi}/ V $, and $F$ is termed \textit{flow} condition, and we explicitly write the time dependence to make clear that this condition should be conserved along the evolution. To gain a better understanding of its physical implications, it is convenient to rewrite the Klein-Gordon equation in terms of $w_{\phi}$ and $\Omega_{\phi}$ \citep{Linder06},
\begin{equation}\label{eq:KleinGordon3}
w'_{\phi} = -3 (1-w_{\phi}^2)\left[ 1-\frac{1}{\sqrt{3F}} \right]
\end{equation}
and by inserting \cref{eq:Flow} in \cref{eq:KleinGordon3} we obtain
\begin{equation}
w'_{\phi} =  \frac{3}{2}(1-w^2_{\phi}) \simeq 3 (1+w_{\phi}),
\end{equation}
which corresponds to the upper limit of thawing evolution in the configuration space \citep{Caldwell05}.\\ 
In the case of a freezing evolution the proof that a dynamical quantity is conserved is straightforward, it is a consequence of the tracking condition in the matter era, which requires $w' = 0$, thus obtaining \citep{Cahn08},
\begin{equation}\label{eq:Flow2}
F(a) = \frac{1}{3}.
\end{equation}
\indent In fact, the flow condition is the result of high friction that any field experiences because of a dominant background component, and it is a physical condition which reflects the nature of evolution in the matter dominant era, thus it should be a natural prior on the initial conditions of the field in Monte Carlo simulations \citep{Cortes10}. Without using this prescription, evolutionary paths of a dark energy fluid can encompass most of the configuration space beyond thawing and freezing priors \citep{Huterer06}. As a matter of fact, the asymptotic value of $F(a)$ at the early-times (hence $F_0$) for the thawing case with $F_0<4/27$ corresponds to a field that has been super-accelerating ($\ddot{\phi}>H\dot{\phi}$), thus breaking matter-domination due to high values of $w_{\phi}$ and $\Omega_{\phi}$ at early-times. Conversely when $F_0>4/27$ the field remains a sub-dominant component  ($\ddot{\phi}<H\dot{\phi}$) with respect to dark matter which dominates the expansion. Analogously in the freezing case, setting $F_0>1/3$ corresponds to super-deceleration, where the potential is extremely flat, while $F_0<1/3$ results in a non-freezing evolution \citep{Cortes10}.\\
Flow condition holds even up to late-times, e.g. $z \sim 2$, before dark energy becomes dominant in the Universe. In \cite{Cortes10,Linder17} an ansatz on the approximate functional form of $F(a)$ at late-times is proposed, which allows to tackle thawing and freezing classes at once, without the need to specify the potential $V$. In the current work we do not follow this approach, but we evaluate the numerical solution of \cref{eq:KleinGordon2} for specific models. In our framework the flow condition appears as an additional prior only at the initial integration time $z_i$, on a combination of free parameters through \cref{eq:Flow,eq:Flow2}, which we detailed in \cref{sec:Methods}.

\section{Dataset}
\label{sec:Data}
\subsection{Baryon acoustic oscillations}
\label{subsec:BAO}
The more recent improvements in galaxy clustering observations \citep{Eisenstein05} provide a wonderful agreement between the early-time CMB constraints and late-time background evolution. The BAO measurements are however obtained only relative to the scale of sound horizon evaluated at the drag epoch $a_d = 1/(1+z_d)$
\begin{equation}\label{eq:sound}
r_{s}(a_d)= \frac{c}{H_0} \int_{0}^{a_d} \frac{ da}{a^2 E(a) \sqrt{1+ a \frac{3 \Omega_b h^2}{4\Omega_{\gamma} h^2}}},
\end{equation}
hence providing a \emph{calibrated} standard ruler which can be utilised to infer cosmological parameters. In the above relation $\Omega_b$ and $\Omega_{\gamma}$ are the densities of baryons and photons today, and $h \equiv H_0/100$. To evaluate \cref{eq:sound} we implement the fitting formula provided in \cite{Aubourg14}, which has been validated for a wide range of models, including dynamical dark energy.\\
In our study we used the estimates of the comoving angular diameter distance $D_M(z)$ and the Hubble rate $H(z)$ provided at $z = \{0.38, 0.51, 0.61\}$ by  \cite{Alam16}, which combines the analysis of different companion works on SDSS DR-12, as a consensus result. At intermediate redshifts we utilised the more recent measurements provided by SDSS-IV eBOSS data release \citep{Zhao18_dr14}, at redshifts $z=\{0.98,1.230,1.526, 1.944\}$. Finally the farthest measurements in redshift are provided by the auto-correlation of the Ly-$\alpha$ forest at $z=2.3$ \citep{Bautista17} and the cross-correlation of Ly-$\alpha$ and quasars at $z=2.4$ \citep{Bourboux17}.

\subsection{Cosmic Microwave Background}
\label{subsec:CMB}
As the discussion in the current work concerns only with the dynamics of the background evolution, and as quintessence is notably known to weakly affect clustering on the linear scales \citep{Alimi09}, in the case of CMB we utilise only the distance-related measurements reported in \cite{Planck15_DE} (hereafter P16) in the form of a compressed likelihood. For this purpose they provide the shift parameter
\begin{equation}\label{eq:R}
R = \sqrt{\Omega_{m0}H^2_0} \frac{D_{M}(z_*)}{c},
\end{equation}
the angular scale of the sound horizon at the last scattering
\begin{equation}\label{eq:l}
l = \pi \frac{D_{M}(z_*)}{r_s(z_*)},
\end{equation}
and the prior on the energy density of baryons $\Omega_b h^2$. Here, $z_*$ is the redshift of decoupling, which is very well approximated by the fitting formula provided in \cite{HuSugiyama95}, with a very weak dependence on cosmological parameters. In our analysis the energy density of radiation $\Omega_{r} h^2$ is fixed by the temperature of CMB, and we also fix the number of effective neutrinos to $3.046$. This likelihood is suited to describe dark energy models with sound speed $c_s \lesssim 1$, which is the case for quintessence, where $c_s = 1$.

\subsection{Supernovae-Ia}
\label{subsec:SN}
The most recent compilation of the SN observations in \cite{Scolnic17} has greatly improved the precision of the observed apparent distance moduli to the extent of having systematic uncertainties which are smaller than those of the statistical nature. Conveniently, these distance  measurements were utilised to calibrate $1/E(z)$ observations at $6$ redshifts, $z=\{0.07,0.2,0.35,0.55,0.9,1.5\}$ in \cite{Riess17_pantheon}. These compressed values are calibrated for a flat Universe and were shown to provide invariant constrains on cosmological parameters \citep{Riess17_pantheon}, even when extended to dynamical dark energy using Chevallier-Michel-Polarski (CPL) parametrization \citep{Chevallier00,Linder02}. Owing to the computational ease and that we are only interested in the flat scenarios, we utilised these 6 measurements along with their covariances.

\subsection{Cosmic Clocks}
\label{subsec:OHD}
Differential dating of galaxies was proposed as a means to estimate the Hubble rate in a cosmology-independent way \citep{Jimenez01}. These estimates do rely on synthetic spectra of simple stellar populations and on models of stellar evolutions. Recently a very robust characterization of the differential aging has been tested \citep{Moresco16}, and it provides an estimate of $H(z)$ at $6 \%$ accuracy (see also \cite{Moresco18} for a recent review on the framework of differential dating). In this work we adopted the measurements provided by \cite{Simon05, Moresco12, Moresco15, Moresco16, Ratsimbazafy17}, listed in table $2$ of \cite{Haridasu18_GP}, which comprises $31$ measurements of $H(z)$ over the redshift interval $z \in (0.0798,1.965)$.\\
In addition to the above mentioned datasets, we have also included the local measurement of the expansion rate $H_0 = (73.48 \pm 1.66)$ \text{km/s Mpc$^{-1} $} \cite{Riess18_H0} (hereafter R18) in our analysis. In the later sections we refer to the SN+CC+BAO dataset combination as low-z. We consider the low-z+P16 combination as the \textit{baseline} for cosmological constraints. Moreover in order to be as model independent from early Universe assumptions as possible, when we utilise low-redshift Universe measurements, namely the low-z or low-z+R18 dataset, we treat the sound horizon $r_s(a_d)$ \cref{eq:sound} as a free parameter \citep{Linder08,Aubourg14,Verde16,Haridasu17_bao}.

\section{Methods}
\label{sec:Methods}
In order to test the quintessence models against data we would need to solve Klein-Gordon \cref{eq:KleinGordon} coupled to the Hubble rate \cref{eq:Hubble}. We instead adopted \cref{eq:KleinGordon2}, \cref{eq:wFun}, \cref{eq:OmegaFun} and \cref{eq:AB}, which enables us to solve for the scalar field evolution as a function of $\ln{(a)}$, thus avoiding to specify the background expansion explicitly. In this formalism for the potentials shown in \cref{tab:potentials} the evolution of the field is determined by the following set of parameters:
\begin{equation}\label{eq:ParamSpace}
\{ w_i , \Omega_i , \mu \}
\end{equation}
where $\mu$ represents the additional free parameter which characterizes the potential. For instance it corresponds to the shift symmetry scale $f$ in the case of PNGB, while DEXP is the only potential where two scalar field free parameters are considered ($\lambda_2$ and $\mu$ in \cref{eq:DEXP}), for which we implement uniform priors. The index $i$ refers to the initial integration time and we used the notation $w_i \equiv w_{\phi}(a_i)$ and $\Omega_i \equiv \Omega_{\phi}(a_i)$.\\
Given this set of parameters, and by the means of \textit{flow} condition (\cref{eq:Flow} or \cref{eq:Flow2}) we can solve for the slope of the potential $\lambda$. In turn this slope depends only on a combination of $\phi_i$ and the parameter $\mu$, thus it can be solved for the initial configuration $\phi_i$ of the field in terms of $\mu$. On the contrary, the initial kinetic term $\phi'_i$ is determined solely by the means of  \cref{eq:AB}, i.e. by $w_i , \Omega_i$. The use of the \textit{flow} condition as a prior at $z_i$ provides us with all the necessary conditions to solve the initial value problem.\\

To make statistical inference for the models, we implement the above formalism in a Bayesian analysis by specifying the likelihood function, which as customary we choose to be a multivariate Gaussian,
\begin{equation}\label{eq:likelihood}
\mathcal{L}(\textbf{y}|{\Theta}) = \exp{\left( -\frac{1}{2} \delta \textbf{y}^{T} \textbf{C}^{-1} \delta \textbf{y}  \right)},
\end{equation}
where $\textbf{C}^{-1}$ is the inverse of the noise covariance matrix associated with data vector $\textbf{y}_{obs}$, while we define the residual vector as,
\begin{equation}
\delta \textbf{y} = \textbf{y}({\Theta})  - \textbf{y}_{obs},
\end{equation}
where $ \textbf{y}({\Theta})$ are theoretical expectations which depend explicitly on the set of parameters $\Theta$. The likelihood function in \cref{eq:likelihood} represents the joint likelihood of all the observations listed in \cref{sec:Data}, where the block-diagonal matrix $\textbf{C}^{-1}$ contains the inverse covariance matrices for each of the independent experiments, and $\delta \textbf{y}$ is a vector of residuals. We update the likelihood with a prior distribution $\mathcal{P}({\Theta})$ on the parameters as,
\begin{equation}
\mathcal{P}({\Theta}|\textbf{y}) = 	\mathcal{L}(\textbf{y}|{\Theta}) \mathcal{P}({\Theta}),
\end{equation}
which explicitly corresponds to the target distribution
\begin{align}\label{eq:posterior}
\mathcal{P}( w_i, \Omega_i, \mu, h , \Omega_b h^2 |\textbf{y}_{obs}) & = \mathcal{P}( w_i, \Omega_i, \mu )  \mathcal{P}( h , \Omega_b h^2)  \notag \\  & \times \mathcal{L}( \textbf{y}_{obs} | D(\Omega_i, w_i , \mu) , h , \Omega_b h^2 ).
\end{align}
\indent In the arguments of the likelihood we defined the field evolution operator $D$, that for a specified set of parameters $\{\Omega_i, w_i, \mu \} $ provides $\Omega_{\phi 0}$ and $f(a)$, which are then used in the likelihood to evaluate the Hubble rate. Secondly, the prior distribution on cosmological parameters $\{h,\Omega_b h^2\}$ has been assumed to be independent, as they do not contribute explicitly to the scalar field evolution. Finally, the \textit{flow} prior at the initial time is easily implemented in the current Bayesian framework as,
\begin{equation}\label{eq:flow_prior}
\mathcal{P}( w_i, \Omega_i, \mu ) = \delta_{D}\left(\alpha - \frac{1+w_i}{\Omega_i \lambda^2(\mu) }\right),
\end{equation}
where $\delta_{D}$ is a Dirac delta function, $\alpha$ is a constant which depends on the class of models, being equal to $4/27$, $1/3$ for thawing and freezing models, respectively. We would also like to remind that there is an additional prior implicitly included in \cref{eq:posterior}, namely the cosmic sum rule $\Omega_{r}(a)+\Omega_{m}(a)+\Omega_{\phi}(a) = 1$.\\
While this scheme is incidentally simple, there are a few caveats one should take care of in the analysis. The selection of the initial redshift $z_i$, whence the field evolution is integrated from, must take into account that (1) the flow condition holds along the matter dominated era, and that (2) CMB data are included in the analysis. While the correct choice would be $z_i = z_*$, it could lead to numerical instabilities due to the smallness of $1+w_i,\Omega_i$ at such an high redshift. In all practical cases of interest, possible early dark energy effects at high $z_i$ are negligible, in fact the contribution of early dark energy can be at most one hundredth of the critical density after recombination \citep{Planck15_DE}. Hence we fix $w_{\phi}(a)=w_i$ for $a<a_i$, and double check that no bias is introduced by comparison with the case where CMB data are not included. As the primary leverage on the constraints is provided by low-redshift datasets, the omission of the CMB distance priors simply provides less-tighter, but similar order of constraints. In this regard we do not find any significant variations in the final inferences in the range $z_i \gtrsim 10$, so we proceed by setting $z_i=50$ for thawing and $z_i=250$ for freezing, which ensures the validity of our approximation. This choice of $z_i$ is indeed deep enough in matter era to ensure no bias in the possible dynamics of the field \citep{Cortes10}. Since thawing models start the evolution particularly degenerate with $\Lambda$CDM, we take a lower $z_i$ than freezings', with the prior range on the EoS of $\log_{10}{(1+w_i)}>-9$.\\
In the case of tracking freezing models we also include the tracking prior on the initial EoS,
\begin{equation}\label{eq:TrackingPrior}
\mathcal{P}(w_i) = \delta_{D}\left( w_i -  \frac{-2(\Gamma-1)}{1+2(\Gamma-1)} \right),
\end{equation}
which is required in order to be consistent with the flow. For example, in the case of IPL or SUGRA models this prior corresponds to a one to one relation with the potential parameter $n$ as,
\begin{equation}\label{eq:trackingIPL}
n = -\frac{2}{w_{i}} -2.
\end{equation}
We also test the case where $n$ is considered a free parameter. While the posterior of $w_i$ is unconstrained, $w_{\phi}$ always tracks the value predicted by the tracking condition \cref{eq:TrackingPrior}, and the posterior of derived parameters converges to the same distribution of the case where the tracking prior is applied, but at the cost of longer MCMC chains. Having affirmed this, we always implement the tracking prior for freezing models. Similarly, EXP has no additional free parameter of the potential to be sampled over, since $\phi_i$ is degenerate with the normalization $V_0$ and is chosen arbitrarily, while $\lambda$ is determined solely by the relation in \cref{eq:Flow}. Conversely, PNGB and PL have a free parameter of the potential to be considered, which determines the initial configuration of the field by solving for $\phi_i$ in \cref{eq:Flow}.\\
\\
We implement the \textit{flow} condition also to the DEXP potential, whose dynamical evolution differs from that of IPL, as DEXP scales with the background EoS ($w_{\phi} = w_{bg} = 0$) and its energy density remains approximately constant in this regime according to \cref{eq:scaling}. The DEXP potential has $3$ parameters associated to $V(\phi)$, which are $ \{ \lambda_1, \lambda_2,\mu, \}$. However $\lambda_1$ is once again determined by \textit{flow} condition, so that scaling condition \cref{eq:scaling} is automatically satisfied initially, and the slope is determined according to $\lambda_1 =\sqrt{3 / \Omega_i}$. In this case we also set a flat prior provided by big bang nucleosynthesis constraints on $\lambda_1>9.4$ \citep{Bean01}, though we see it is irrelevant for the posterior. In DEXP potential the initial configuration of the field is degenerate with $V_{0}$ and $\mu$, thus we can set $\phi_i=0$. This potential is also expected to behave like a thawer under the influence of non-zero $\lambda_2$ at late-times, as in the case of EXP model.
\\
We use the publicly available \texttt{emcee}\footnote{\href{http://dfm.io/emcee/current/}{http://dfm.io/emcee/current/}} \citep{Foreman-Mackey13} code, which implements an affine invariant Metropolis-Hastings sampler to ease the exploration of the posterior distribution. For the analysis of the chains and creating plots we utilise the \texttt{getdist}\footnote{\href{https://getdist.readthedocs.io/}{https://getdist.readthedocs.io/}} package  .


\section{Results}
\label{sec:Results}
%
\begin{figure}
	\includegraphics[width=0.5\textwidth]{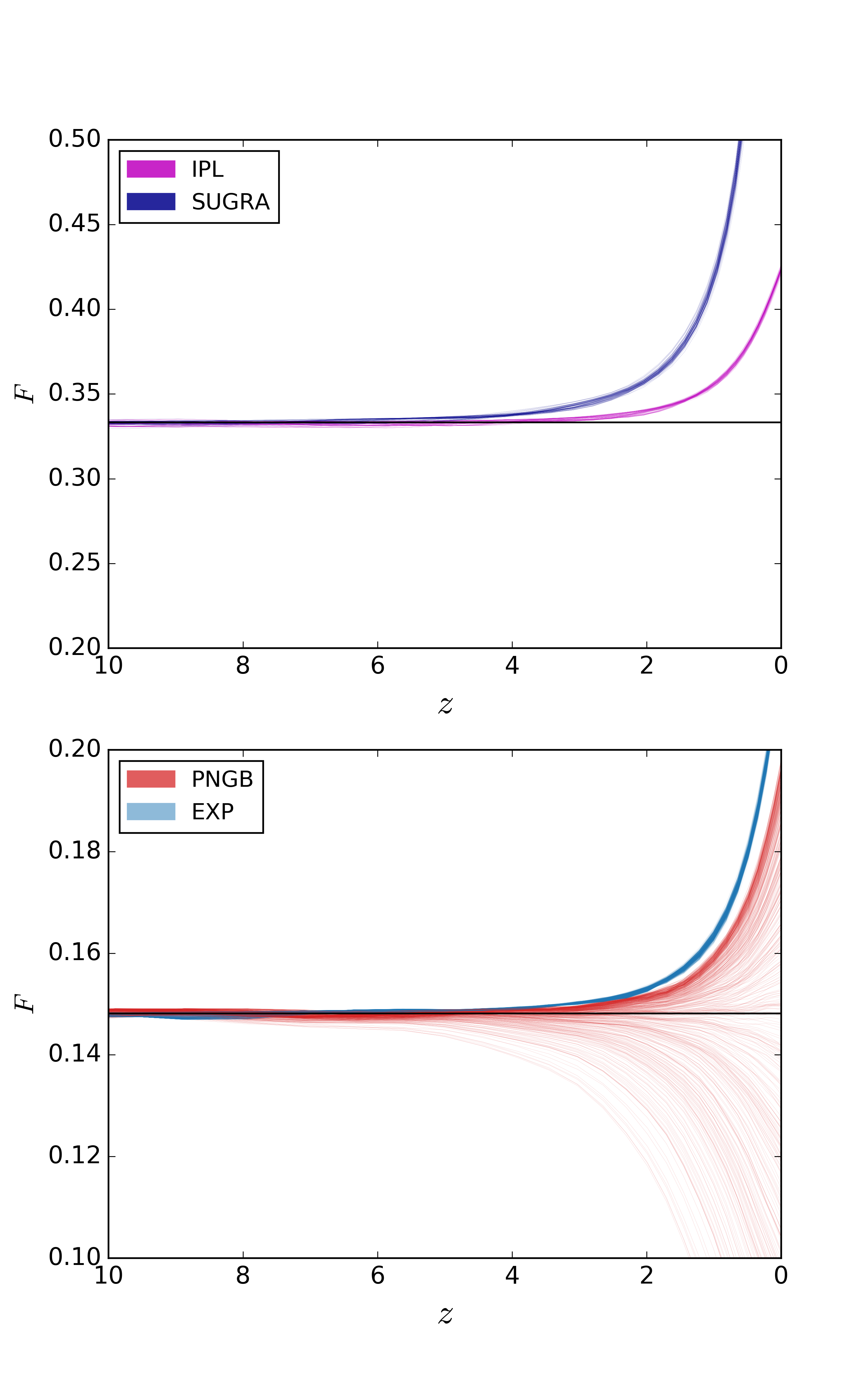}
	\caption{ Evolution of the \textit{flow} term for a few randomly drawn samples from the posterior distribution, obtained using the low-z+R18+P16 dataset. Both in the case of freezing (top panel) and thawing (bottom panel) the asymptotic \textit{flow} ($F_0$) value is conserved along the matter era, and it departs around $z\sim 3$. In contrast to the other models, the additional degree of freedom ($f$) in the PNGB model enables a more varied evolution of $F(a)$ at late-time (details in the text).}
	\label{fig:flow}
\end{figure}
As mentioned in the earlier section, the \textit{flow} prior is fixed at the initial redshift of integration according to \cref{eq:flow_prior}. To begin with we verify whether this condition is supported by data at successive times by inspecting the inferred evolution of $F(a)$. In  \cref{fig:flow} we show a few trajectories drawn randomly from the MCMCs constrained by the dataset low-z+R18+P16. We confirm that the flow condition is indeed conserved along all the matter-dominated era, and is supported by data until before the contribution of dark energy to the expansion rate becomes relevant. The fact that $F(a)$ is conserved along the evolution during the matter dominated era, in turn is a cross-validation for the effectiveness of the formalism, and of the dynamical stability of the \textit{flow} prior. The late-time evolution of the $F(a)$ depends on the potential, e.g. in the case of thawing PNGB it depends on shift symmetry scale. As is shown in the lower panel of \cref{fig:flow}, it is evident that for a certain combination of lower $f$ and higher $\Omega_{\phi}$ values, we have a decrease of $F(a)$ from the asymptotic value of $4/27$ \cite{Cahn08}. It is also worth noting that this departure accordingly starts earlier, asserting an evolution of the super-acceleration kind, as is also elaborated in \cref{sec:Theory2}. All the other potentials retain a strong correlation between $w_{\phi}(a)$, $\Omega_{\phi}(a)$ and $F(a)$, until the present time.\\
$F(a)$ for PL model is not shown as it coincides with EXP, while for DEXP model the \textit{flow} is conserved only deep in the matter era, but it starts to slowly depart from it at early time $z\gtrsim 50$. The contribution of dark energy becomes relevant for the dynamics much earlier than tracking freezings and thawings, since for this model we find that $\Omega_i \sim \Omega_{\phi}(a) \sim 10^{-4}$ along the matter era, while for trackers $\Omega_i \sim 10^{-6}$ (starting from the same initial redshift $z_i=250$). Such behaviour is anyway expected in a scaling regime, where an early transition to acceleration can occur \citep{Tsujikawa13}.\\
Henceforth, we systematically compare the inferences on the models, making the distinction between the use of 3 different dataset combinations: the complete dataset low-z+R18+P16, the baseline low-z+P16 and low-z+R18, as elaborated at the end of \cref{subsec:OHD}. The constraints we obtained are summarized in \cref{tab:results,tab:results_noH0,tab:results_low} for each of the three datasets, respectively. Wherein we show the $2 \sigma$ credible intervals on the minimal set of parameters which are sufficient to integrate Klein-Gordon equation, in addition to constraints on $h$ and on the derived parameters, $\{w_0, \Omega_{\phi 0}\}$.\\

\subsection{PNGB \& IPL}
\label{subsec:PNGB_IPL}
\begin{figure}
	\includegraphics[width=0.45\textwidth]{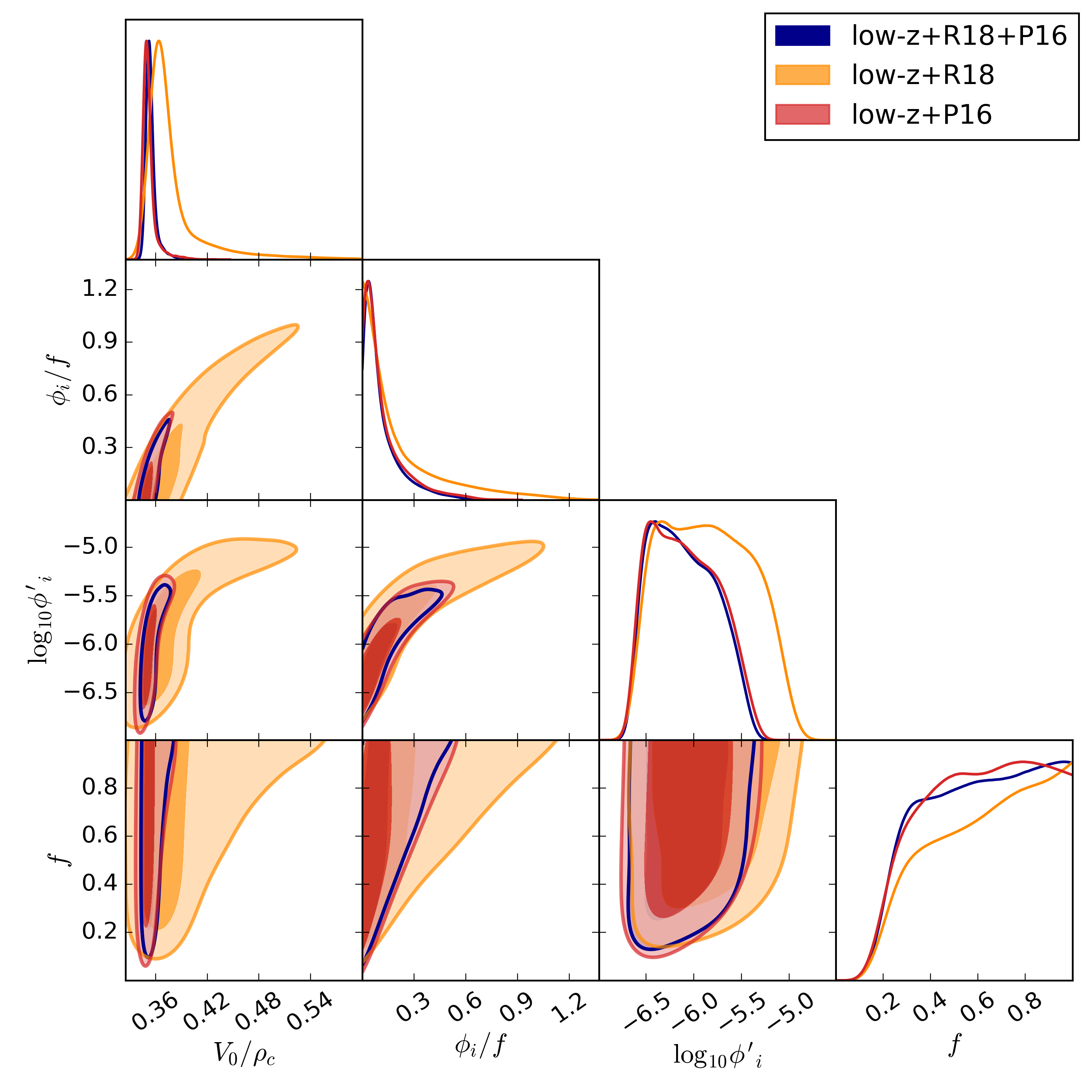}
	\caption{$2 \sigma$ credible intervals for PNGB initial conditions and potential parameters. The effects of different datasets are shown. As expected the major changes are due to the addition of P16 dataset. }
	\label{fig:IC_PNGB}
\end{figure}

In the case of PNGB, we find that the field starting very close to $\phi_{i} \sim 0$ is the most likely scenario, as is also shown in the one-dimensional posterior of $\phi_i$ in \cref{fig:IC_PNGB}. The shift symmetry of PNGB slightly favours larger values of $f$, with $95\%$ of the samples lying in the region $f > 0.24$.\\
Interestingly, the possibility that the field starts rolling from the top of the potential, while is hard to motivate in the standard quantum field scenario \citep{Banks03,Adak14}, it is justified in the framework of string axiverse \citep{Kamionkowski14}. Similar statistical analysis of PNGB were earlier performed in \cite{Abrahamse08,Rubin09,Adak14,Barreto15}. In \cite{Adak14}, considering a parameter space limited to either of the paired combinations from amongst $\{\phi_i, \Omega_{\phi},f\}$, it was concluded that the shift symmetry scale $f < 0.10 \ m_{pl}$ is almost ruled out by observations, and the most likely scenario corresponds to $\phi_i \simeq 0$, near the top of the potential. Similar inference was also made in \cite{Barreto15}, which we agree with at an even higher degree of confidence, obtained using our baseline dataset. However, in contrast to the study by \cite{Barreto15}, where the analysis was performed in an extended parameter space much like the current work, we find a discrepancy in the posterior distribution of $\phi_i/f$ which they predict to be marginally dependent on the choice of the parametrization of $f$ (see figure $3$ and $7$ of their analysis). We cannot immediately address this difference due to the diverse choice of $z_i$, which in their case was set to the higher redshift $z_i = 10^{15}$, since the field dynamics is expected to be frozen at until late times in a thawing scenario.\\
We also find our results to be in agreement with the more recent conclusions of \cite{Emami16}, where the authors explicitly address the string axiverse scenario as a viable phenomenological model. Their analysis was performed with limited dataset and stricter prior ranges on the parameters, and without the additional \textit{flow} condition, which in fact explains the considerable improvement in our results, e.g. by considering their inferences on $w_0,\Omega_{\phi 0}$ shown in figure $2$ of their analysis. We show the posterior distributions of the field parameters in \cref{fig:IC_PNGB}, wherein one can notice that the shift symmetry scale $f$ seems unaffected by the choice of dataset, while the changes are apparent in the initial conditions $\{\phi_i,\phi'_i\}$. These mild differences, however become more evident in the posteriors of derived observable quantities (top panel of \cref{fig:observables}). As expected, we find that the effects of including P16 are twofold, a much tighter constraint on the energy density $\Omega_{\phi 0}$ and a lower value of $w_0$. We anticipate here that this effect is retained by all the quintessence models, in other words the dynamics of the field are severely limited by `high-redshift' P16, as it was also pointed out by \cite{Peirone17, Park18b}.\\
From the posterior of $h \equiv H_0/100$ shown in top panel of \cref{fig:observables}, one can notice that the addition of R18 to low-z+P16 does not affect the inferred value of the current expansion rate, which shows R18 as an outlier at about $\sim 2.5 \sigma$. Clearly, quintessence models prefer a lower value of $H_0$, more in agreement with CMB constraints \citep{Planck15_DE}. As a matter of fact $H_0$ tension can be eased with values of the EoS favouring a phantom scenario, thus explaining the limited dynamics retained by quintessence with the inclusion of P16 \citep{DiValentino17}.\\
In conclusion our findings suggest that a string axiverse PNGB, also addressed more recently in \cite{Cicoli18} as a plausible solution to dark energy, is indeed consistent with observations, however unable to ease $H_0$ tension and strongly degenerate with $\Lambda$CDM \citep{Linder15}. In fact, from the evolution of the EoS of PNGB in \cref{fig:eos}, we see that the initial value of EoS can differ as much as one part over a million from the vacuum EoS expected value, while today the deviation is of the order of $\sim 0.01$.
All the arguments presented so far for the \textit{flow} prior also stand true for the PL potential. However, the evolution of the field under PL potential is not affected by the free parameter $n$, and its evolution is similar to the case of EXP. This indicates that the slope of the potential as constrained by current data is indeed very flat for both cases, as thawers evolve similarly and yield the same phenomenology in the low-redshift range \citep{Scherrer07}. The initial configuration of the field is always fine-tunable for a given $n$ by the means of relation \ref{eq:Flow}, and the posterior of $n$ indeed shows uniform probability in the prior range ($1<n<4$) we considered, at the price of having very large super-Planckian values for the configuration of the field, and a tiny normalization scale $V_0$ with respect to all the other quintessence models. On the basis of technical naturalness alone this model seems to be disfavoured by our constraints.\\
\begin{figure}
	\centering
	\includegraphics[width=0.45\textwidth]{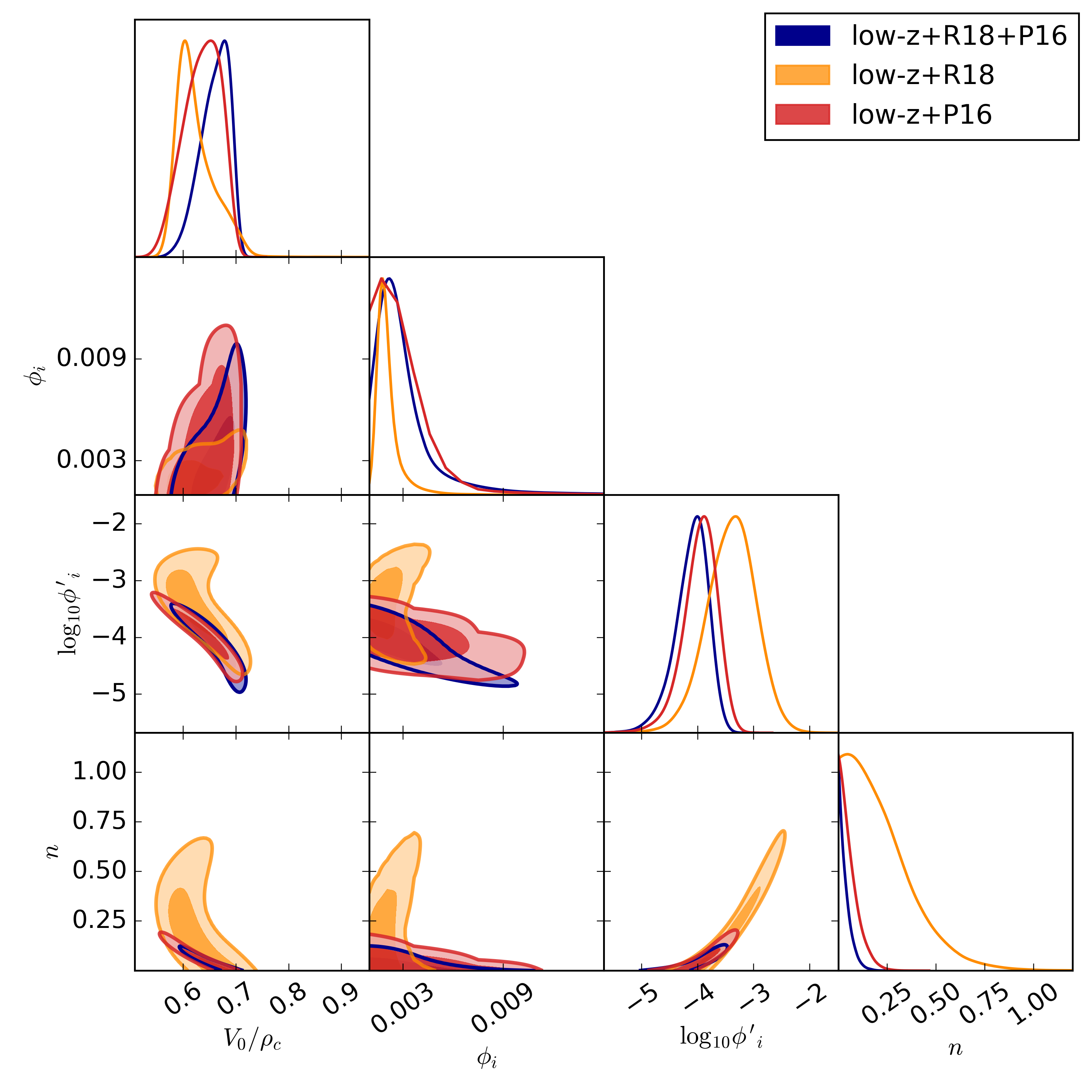}
	\caption{\label{fig:IC_IPL}
		$2 \sigma$ credible intervals for IPL initial conditions and potential parameters. The effects of different datasets are shown. The addition of P16 places a more stringent bound on $n$, namely on the EoS. Analogously R18 prior pushes the EoS more towards $-1$.}
\end{figure}
In the case of freezing, the stability of flow around the theoretical value $1/3$ is equivalent to the validity of the tracking regime in the matter era. At late-times the field approaches a flatter region of the potential, so that $F(a)$ monotonically increases. While SUGRA and IPL lead to very similar evolutions, there is a small distinction due to the fact that SUGRA can have a greater value of EoS in the past and reach a smaller value by the present time, a peculiarity which is well known \citep{Brax99}, and it is shown in figure \ref{fig:eos}, where the freezing phase of SUGRA is more apparent. Nonetheless, the overall evolution of both these potentials is practically coincident, with very similar initial conditions. We notice that IPL and SUGRA are also degenerate with $\Lambda$CDM, though they show more deviation in the past evolution with respect to a thawing scenario. The degree of degeneracy can be estimated by the parameter $n$, which is related to $w_i$ in the matter era owing to \cref{eq:trackingIPL}. For instance, we find the parameters to be bounded by $n<0.10$ ($w_0<-0.97$) at $2 \sigma$ level, when using the complete dataset. It corresponds to $n<0.15$ ($w_0<-0.95$) without R18, while it can be as high as $n<0.50$ ($w_0<-0.87$) in the case of low-z+R18 dataset. This makes it apparent that similar to the case of PNGB, the degree of degeneracy is more severe with the inclusion of P16, the values of the EoS are lower and the $H_0$ tension remains (c.f. bottom panel of \cref{fig:observables}). We notice that our results are consistent with \cite{Park18a,Park18b}, whose datasets are similar to our low-z+P16 and low-z, respectively. While we have slightly improved constraints, our major inferences agree with these works. A similiar progressive change in terms of constraints on $n,w_0$ is found for SUGRA potential as well.\\
An alternate approach to test quintessence models is to adopt approximate analytical solutions of the initial value problem. This approach was implemented by \cite{Chiba12} and more recently the results have been updated in \cite{Durrive18}, where they use the full Planck likelihood of 2015 data release together with a few BAO points and Joint Light-curve Analysis. Surprisingly, they found that thawers can have an EoS value today as high as $w_0=-0.473$ at $2 \sigma$ level, which is in much looser than what we find here, and even less stringent than the similar previous analysis \citep{Chiba12}. Conversely, for freezing models our results are very consistent, as they found $w_0<-0.92$ in comparison to our $w_0<-0.95$ for IPL and $w_0<-0.97$ for SUGRA obtained using low-z+P16 dataset.\\
\begin{figure}
	\centering
	\includegraphics[width=0.5\textwidth]{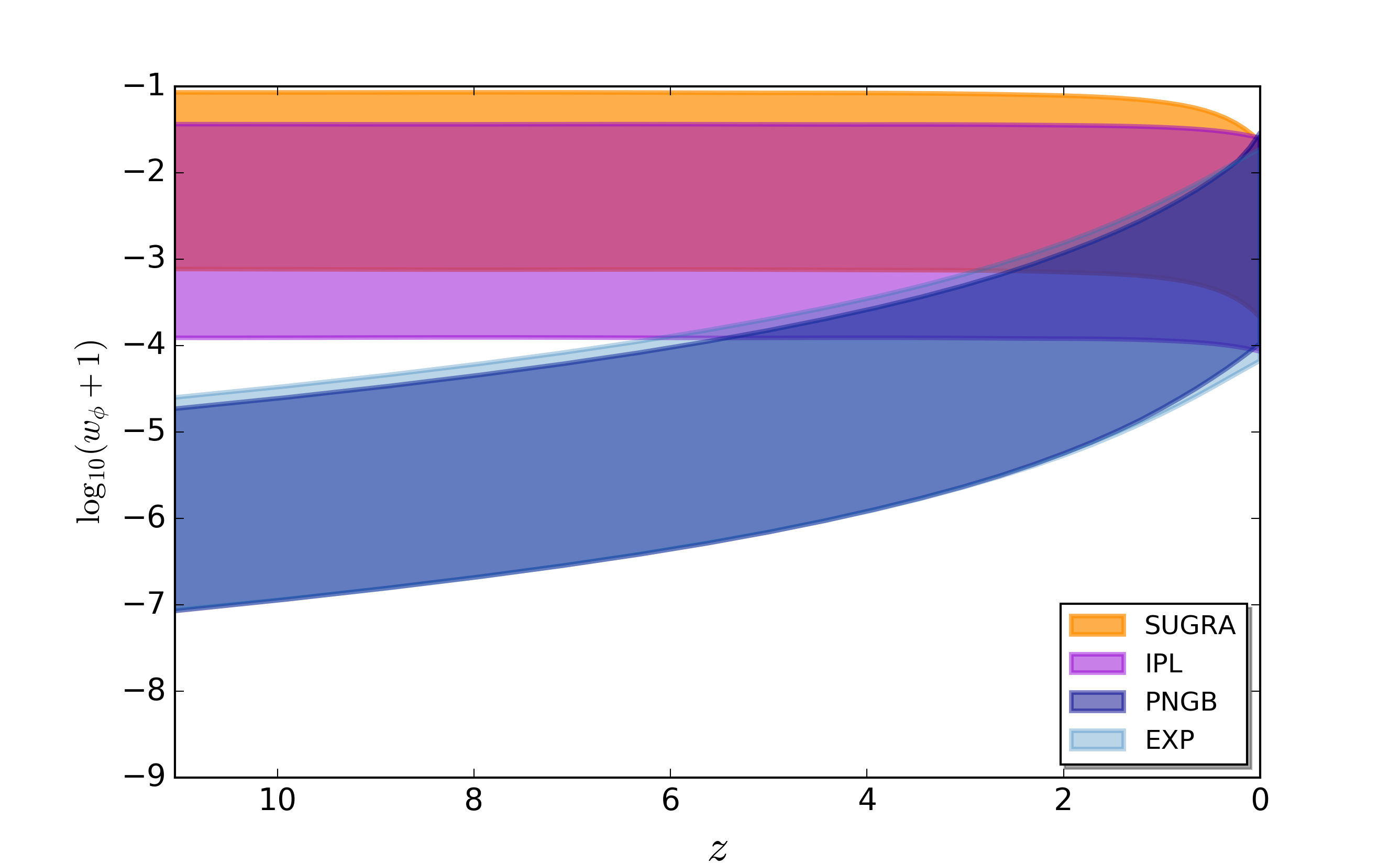}
	\caption{Evolution of EoS consistent with the complete dataset low-z+R18+P16. The shaded regions correspond to the area between $2.5$th and $97.5$th percentiles of the samples of the posterior. As expected all the models reach a similar value of the EoS today, in light of low-redshift data, while their past evolution is constrained by their dynamical nature. Thawers (PNGB,EXP) are more degenerate with $\Lambda$CDM in the matter era due to the high Hubble friction, while freezing models (SUGRA,IPL) track the background EoS ($w_{bg}=0$) thus $w_{\phi}$ is almost a constant during the matter era. The regions shown here are only representative for the nature of the potentials and should not be taken as credible intervals.}
	\label{fig:eos}
\end{figure}
\begin{figure}
	\begin{minipage}[c]{0.45\textwidth}
		\includegraphics[width=\columnwidth]{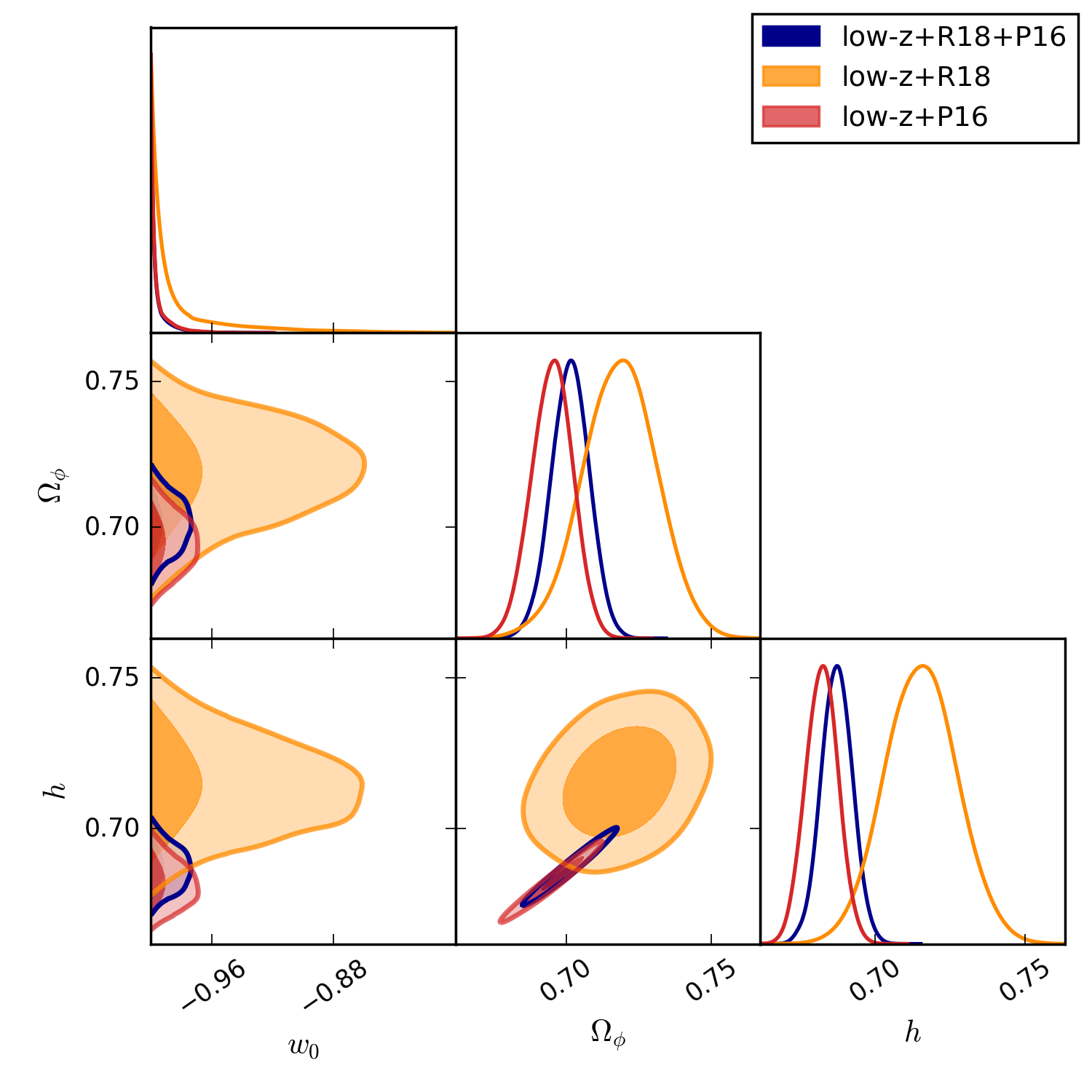}	
	\end{minipage}\\
	\begin{minipage}[c]{0.45\textwidth}
		\includegraphics[width=\columnwidth]{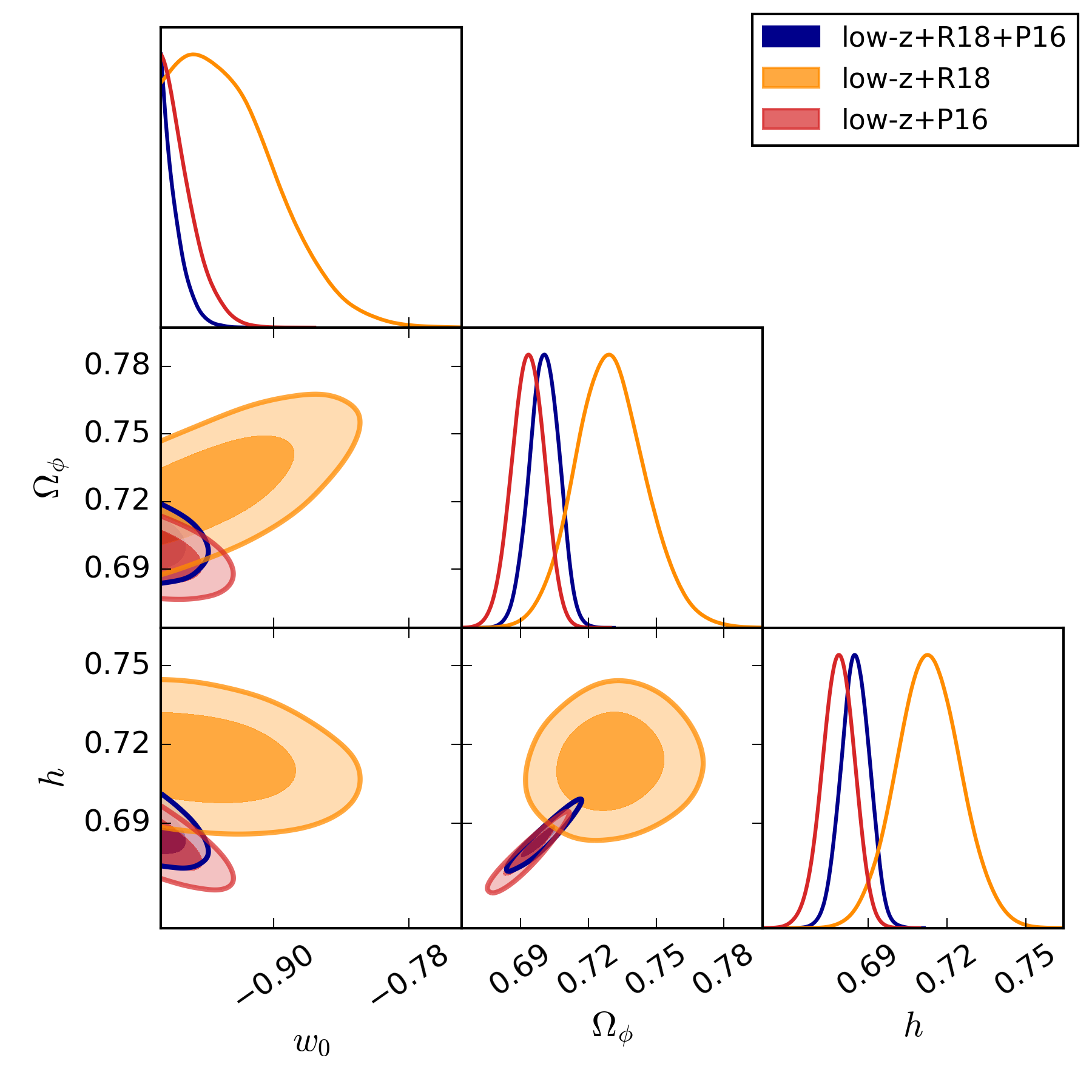}	
	\end{minipage}
	\caption{\label{fig:observables} \textit{Top panel:} $2 \sigma$ credible intervals of PNGB observables, derived from initial conditions of figure \cref{fig:IC_PNGB}. \textit{Bottom panel:} $2 \sigma$ credible intervals of IPL observables, derived from initial conditions of figure \cref{fig:IC_IPL}. In both cases it becomes apparent that the stringiest constraints are placed by P16, and the fact that quintessence models cannot ease the H0 tension with R18 face value.}
\end{figure}
\subsection{EXP, DEXP \& Swampland}
\label{subsec:swampland}
Potentials such as EXP and DEXP are often considered in the context of string theories, which have lately been of interest owing to a newer criterion conjectured on the viability of the de Sitter vacua \citep{Obied18}. Specifically, it was argued that for any consistent theory of gravity that has an UV completion:
\begin{itemize}	
	\item The slope of the potential has to satisfy the limit $\frac{\nabla V}{V} \gtrsim d \sim \mathcal{O}(m_{pl}) $,
\end{itemize}
and failing to satisfy, the theory falls into the swampland. The validity of this conjecture has important phenomenological consequences, i.e. in the context of string theories there would be no place for a cosmological constant, and dark energy is justified only if we consider a dynamical scalar field evolving under the potential $V(\phi)$. In light of this ansatz several works have explored the implications of these criteria for quintessence, both theoretically \citep{Chiang18,Cicoli18,Marsh18} and phenomenologically \citep{Heisenberg18,Agrawal18,Akrami18}. These works have focused on the constraints of the slopes of EXP and DEXP potentials. As we constraint these models also in our analysis, we extend the discussion along the same lines. We verified the stability of the \textit{flow} also in the case of the thawing EXP (see \cref{fig:flow}). This model is completely determined by the specification of $w_i$ and $\Omega_i$ as $\lambda$ is a posteriori derived using \cref{eq:Flow}. We find that the slope of EXP is very flat, as it can be as small as $\lambda < 0.30$ at $95 \%$ confidence level using the complete dataset low-z+R18+P16, which indicates a strong tension with the aforementioned swampland conjecture if we expect $d \sim \mathcal{O}(m_{pl})$. This limit is to be compared against the upper limit of $\lambda \lesssim 0.60$ presented in \cite{Akrami18}, using a very similar dataset with the inclusion of R18 and P16.
The tension remains also in the case of DEXP, with $\lambda_2<0.47$ at $95 \%$ confidence level (see \cref{tab:results}). The inferred range of $\lambda_1$ is determined by the scaling property of the EXP potential $\lambda_1 \sim \Omega_{i}^{-1/2}$, so that a broad upper bound is always found (the lower prior $\lambda_1>9.4$ is not explicited in the tables). The other parameter of the model $\mu=V_2/V_1$, usually assumed to unity in standard analysis \citep{Chiba12,Durrive18,Akrami18}, is constrained by our MCMC study and it corresponds to $\log_{10}{\mu} = -2.6^{+1.3}_{-1.4}$. The smallness of $\mu$ is expected in order to realize the condition $e^{-\lambda_1 \phi} \ggg \mu e^{-\lambda_2 \phi}$, thus ensuring the scaling solution at early times. To the best of our knowledge our study is the first that implements such a comprehensive analysis of DEXP. Indeed in \cite{Chiba12} it was fixed $\mu=1$ and parameters constraints were inferred from the profile of the likelihood in a reduced parameter space, while in \cite{Durrive18} only a fitting parametrization of the EoS in considered in the limit $\lambda_2=0$. \cite{Durrive18} finds that P16 favours an high redshift of transition to acceleration, about $z \gtrsim 10$, which is in agreement with the \textit{flow} condition departing from $1/3$ at early time. We assess that the initial energy density for DEXP is $\Omega_i \sim 10^{-4}$ at $z_i=250$, which is also consistent with P16 model independent upper bound on $\Omega_{de}$ at $z=\{50,200,1000 \}$.\\
The swampland conjecture is also in tension with the baseline dataset for both EXP and DEXP, with $2 \sigma$ upper bounds on the slopes $\lambda<0.31$ and $\lambda<0.54$ respectively. The tension is less severe, though it remains, only in the case of low-redshift data alone.\\

\begin{figure}
	\centering
	\includegraphics[width=0.5\textwidth]{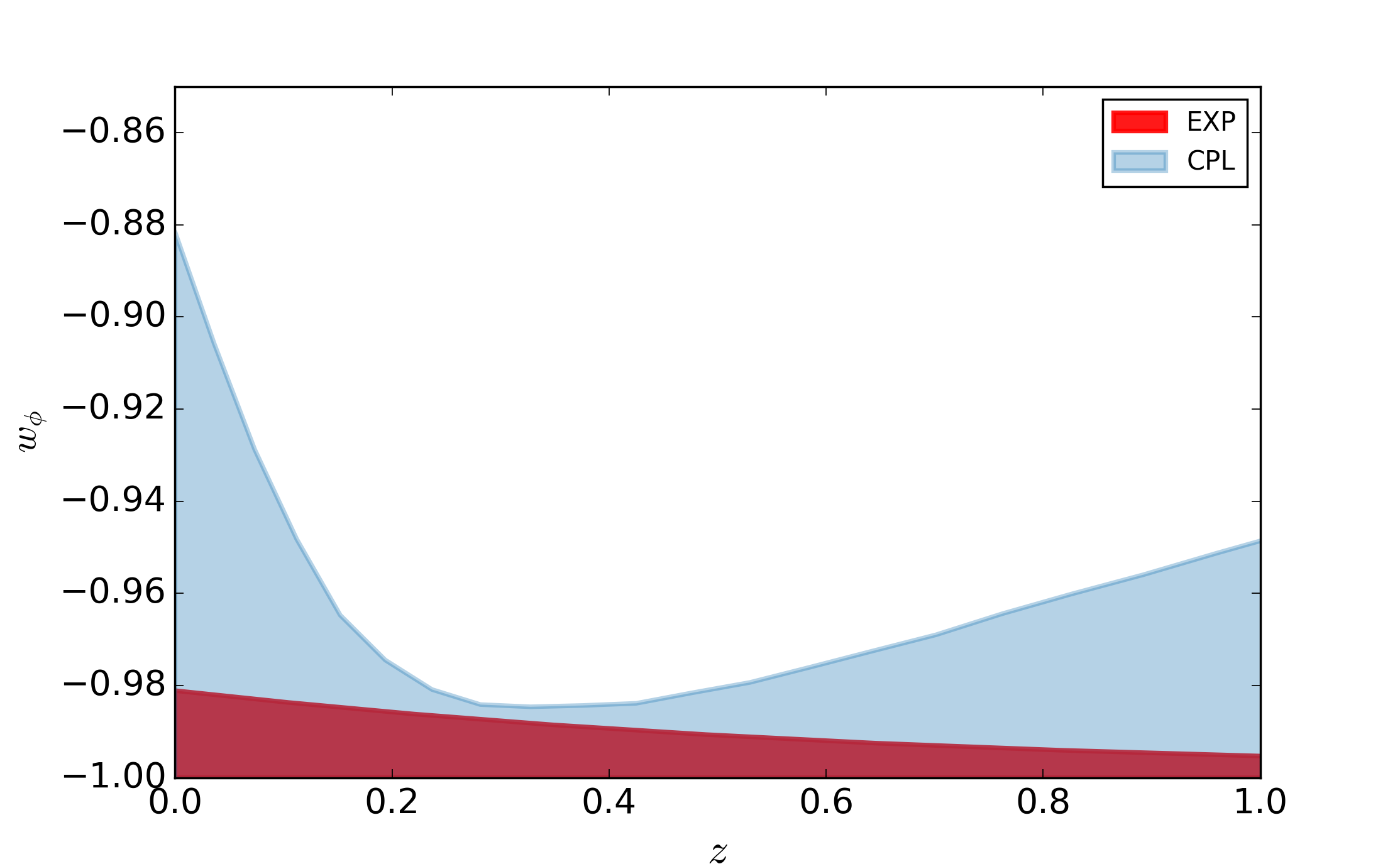}
	\caption{Comparison of the $2 \sigma$ credible intervals for CPL and EXP. At $2 \sigma$ level the slope of EXP is bound to be $\lambda<0.30$ for the low-z+P16+R18 dataset. These limits do not change substantially from our baseline dataset.}
	\label{fig:lambda}
\end{figure}

In our study we obtain improvement in the constraints over the analysis performed in \cite{Agrawal18,Akrami18,Heisenberg18}, which are based on the CPL parametrization of the EoS $\left[w(a) = w_0 + w_a(1-a)\right]$. The improvement essentially resides in the use of a more complete dataset, model-dependent approach and also in the use of the \textit{flow} prior. For a qualitative comparison we show in \cref{fig:lambda} the $2 \sigma$ level constraints on EoS for CPL and EXP potential, which is to be confronted with figure 1 of \cite{Heisenberg18} and figure 5 of \cite{Akrami18}. It can be clearly seen that the previously argued upon upper limit on the value of $\lambda$ is further reduced. Our inference for the CPL parametrization is also in agreement with the recent analysis in \cite{Dai18}, done utilising very similar dataset. In passing we would also like to mention that the mapping between CPL space and the field should however be performed according to the calibrated prescriptions for thawing models \citep{DePutter08}, which could reduce the available region in the  $(w_0,w_a)$ plane corresponding to $\lambda$.
%
\subsection{Model Selection}
\label{subsec:model_selection}
Finally we infer which among these potentials is statistically favoured in our analysis. We achieve this by comparing the deviance information criterion (DIC) \citep{Liddle07,Trotta08} of each model, defined as
\begin{equation}\label{eq:DIC}
	 \textrm{DIC} = \chi^2_{eff}(\hat{\Theta}) + 2 p_D,
\end{equation}
where we defined $\chi^2_{eff} \equiv -2 \ln{\mathcal{L}}$, and $\hat{\Theta}$ refers to the model parameters at the best fit value of the likelihood. The last term on the right-hand side is the model complexity $p_D \equiv \overline{\chi^2_{eff}} - \chi^2_{eff}(\hat{\Theta})$, with the overline indicating the average over the ensemble. $p_D$ is an estimate of the number of \textit{effective} parameters which are constrained by the data \citep{Kunz06}, which in \cref{eq:DIC} is combined with the goodness of the fit $\chi^2_{eff}(\hat{\Theta})$, so that DIC does not penalize a model on the basis of number of free parameters alone, but based on the constrained ones. Other criteria such as Akaike information criterion (AIC) \citep{Akaike74} and its corrected version $\textrm{AIC}_c$, previously adopted by e.g. \cite{Verde10,Shafer15,Haridasu17_bao}, are based on the value of the likelihood at best-fit, on the number of free parameters of the model and on the number of data points, thus they do penalize substantially as we add more free parameters.\\ 

As is customary, the evidence of a model $x$ is quoted with respect to a reference model ($\Lambda$CDM) in terms of $\Delta \textrm{DIC}_x = \textrm{DIC}_{x}-\textrm{DIC}_{\Lambda \textrm{CDM}}$, $\Delta \textrm{AIC}_c = \textrm{AIC}_{c,x}-\textrm{AIC}_{c,\Lambda \textrm{CDM}}$, and information criteria are gauged by the Jeffreys scale \citep{Jeffreys61}, where $\Delta \in \{-2,-5\}$ indicates moderate/positive evidence and $\Delta \in \{-5,-10\}$ implies a strong evidence \citep{Trotta08,Grandis16}. The results are shown in the last two rows of \cref{tab:results,tab:results_low,tab:results_noH0}, for a comparison of both AIC and DIC, respectively. As expected we find the corrected Akaike information criterion to penalise all the models heavily and moderately disfavouring the quintessence models, while DIC shows that all the thawing models are statistically equivalent, with the exception of freezing IPL and SUGRA which are moderately disfavoured. When a low-redshift analysis is performed all the models are statistically consistent, but in the case P16 is added to the dataset, thawers become the favoured ones. The addition of R18 does penalize freezing models even more, while thawing kind remain consistent with $\Lambda$CDM. These results highlight the need to adopt appropriate criteria to perform model selection. In this regard we find discrepant results with respect to previous AIC-based detection of dynamical dark energy for IPL in \cite{Sola17b,Sola17a,Sola17}, where they have utilised the large scale structure (LSS) dataset, which is not considered in the current work. However, in \cite{Park18a} using a very similar dataset (including LSS) as \cite{Sola17b}, no such preference for IPL was found, which our results are in agreement with. In a more recent work \cite{Sola18} the same authors finds consistent results with the current analysis and \cite{Park18a}, thus settling these differences. Interestingly, they find that the addition of LSS data extracted from the bispectrum \cite{Gil-Marin16} do provide evidence of dynamics for IPL. A deviation from general relativity due to the bispectrum was also found in \cite{Gil-Marin16}.

From our analysis we conclude that quintessence models, where a single scalar field drives the late time accelerated expansion of the Universe, are still viable solution. While there is no evidence in favour of dynamics, there is also no evidence against dynamics of quintessence kind, as opposed to the claim disfavouring quintessence \citep{Zhao17_nature}. More recently, \cite{Roy18} have commented on the nature of the arbitrariness in the choices for quintessence potentials and claimed they are indistinguishable, since current data can constrain neither the potential parameters nor initial conditions (what they call active and passive parameters respectively). Their study is based on parametrizing $V(\phi)$, similarly to the approach of \cite{Huterer06}. It was already argued in \cite{Cortes10} that an analysis based on the expansion of the potential may lead to biases in the dynamics of the field, while the flow-based approach can prevent the same, and enables us to find that current data can constrain potential parameters and initial conditions for quintessence fields. In fact we confirm that there is no shortage of thawers and that models are actually distinguishable, since the relative $\Delta \textrm{DIC}$ between quintessence models shows thawing potentials to fare better. In addition to the already known, preferred naturalness of the thawing models \citep{Linder15}, we show that they are also moderately favoured over freezing kind. Finally, we also note that in \cite{Peirone17}, assuming two categories of priors, one with strict mathematical priors and a second one with an additional physical ($w_0>-1$) prior, it was shown that the CPL model is favoured with $\Delta \textrm{DIC} \sim - 5.4$ only when $w_0<-1$ parameter space is allowed, on the contrary when the physical prior is imposed, it is disfavoured with $\Delta \textrm{DIC} \sim 4.6 $. In our analysis however, using the same information criteria and analogous prior ($w(z)>-1$) we do not find evidence of such significance, with an utmost disadvantage for the freezing models at $\Delta \textrm{DIC} \sim 2.7$. While there is clearly a difference in the data utilised in earlier work and this work, the difference could also arise from the fact that the same parameter space tested in \cite{Peirone17}, which in our analysis is further reduced, and in turn is explicitly split into two separate regions for the freezing and thawing nature of the potentials.

\section{Summary \& Conclusions}
\label{sec:Conclusions}
In this study we implemented a statistical framework to place constraints on quintessence models of dark energy. There is a conserved non-linear combination of physical quantities in the long matter era, the flow condition (\cref{eq:Flow,eq:Flow2} and \cref{fig:flow}), which reflects the nature of the evolution of the field, so that it can be used to fine-tune Monte Carlo simulations. We implemented it in a Bayesian formalism as a prior on the initial conditions and on the parameters of the potentials (\cref{eq:flow_prior}). We performed a MCMC exploration of the target distribution \cref{eq:posterior} of the models listed in \cref{tab:potentials}. In this Bayesian framework, quintessence evolutionary paths are constrained based on a compilation of low and high redshift experiments, such as BAO, SN, CC, R18 and CMB. The resulting posterior corresponds to all cosmic histories which are consistent with present observations, in a Universe where acceleration is powered by quintessence. Thanks to this approach we were able to set constraints on the scalar field initial conditions and parameters of the potentials, in the most extended parameter space consistent with these models, a novelty with respect to previous analysis (see \cref{fig:IC_PNGB,fig:IC_IPL}). Our results are summarized in \cref{tab:results,tab:results_low,tab:results_noH0}, where we provide the $2 \sigma$ credible intervals on the set of parameters required to integrate the initial value problem for the system of quintessence evolution (\cref{eq:Hubble,eq:KleinGordon2}).  Here we summarize the key improvements and major results from our analysis,
\begin{itemize}
    
	\item We implement a more physical prior (\textit{flow}) on the initial conditions on the field, in order to constrain the potential parameters, and utilising the most recent data, we update the existing constraints. At $2\sigma$ we find stricter bounds on the potential parameters such as, $f/m_{pl}>0.263$ for the pNGB and $n<0.152$ in the case of IPL. For the DEXP aside form $\lambda_1, \lambda_2$ parameters, we also constraint $\log_{10}{\mu} \sim -2.5^{+1.3}_{-1.5}$, ratio of the two potentials. 
        \newline
    \item We infer that the current data do not provide any evidence against thawing models, which are statistically equivalent to $\Lambda$CDM. However, $\Lambda$CDM seems to be moderately preferred over the freezing kind with $\Delta$DIC $\sim 2.7$. This preference in fact increases to $\Delta$DIC $\sim 3.3$ using low-z+P16+R18. 

	\item While it is already known that the quintessence models are unable to ease the $H_0$ tension, we reaffirm by showing that R18 remains an outlier at $\sim 2.5 \sigma$, when included in the analysis. In fact, only from our low-z+R18 analysis, which is independent of early-times physics, we find that an higher value of $H_0$ can be facilitated with corresponding increase in $\Omega_{\phi0}$
    
	\item We report an increased tension on swampland criterion (see \cref{subsec:swampland}), with the estimate $\lambda \lesssim 0.31$ at $2 \sigma$ for the EXP, and $\lambda \lesssim 0.54$ for the DEXP. When using the dataset low-z+R18, this stringent constraint is relaxed with $\lambda < 0.871$ for DEXP.
\end{itemize}
While the results present in here are not strictly conclusive in nature, the road ahead to unravel the nature of late-time acceleration is long, and there is a need to implement better analysis to perform the model selection. Quintessence models, being the first to be proposed, are yet to be decisively favoured/disfavoured by the observations.

\section*{Acknowledgements}
Authors would like to acknowledge financial support by ASI Grant No. 2016-24-H.0.

\bibliographystyle{apsrev4-1}
\bibliography{bibliografia2}

\appendix*
\section{Tables of results}
{\renewcommand{\arraystretch}{2.}%
	We present the 2σ credible intervals on the minimal set of parameters that are sufficient to integrate the Klein-Gordon \cref{eq:KleinGordon}, in addition to constraints on h and on the derived parameters $\{ w_0,\Omega_{\phi0} \}$. We quote the inferences for all the three datasets combinations considered in the text, the complete low-z + R18 + P16 and low-z + R18 data only. In all cases we quantify the statistical significance of the results by the means of information criteria AIC and DIC with respect to $\Lambda$CDM.
	\begin{table*}
		\caption{We show the $2\sigma$ confidence limits of the relevant parameters for all the potentials, obtained using the low-z+R18+P16 datasets. Index $n$ in the case of PL is unconstrained. $\lambda$ refers to $\lambda_2$ in the case of DEXP. The parameter $\lambda_1$ has flat prior $\lambda_2>9.4$, while $f$ has flat prior $f < 1$.}
		\label{tab:results}
		\begin{tabular} { l  c c c c c c }
			\hline
			&  PNGB  & EXP  & PL & IPL & SUGRA & DEXP \\ 
			\hline
			{\boldmath$w_0            $} & $< -0.981                $  &  $< -0.986                  $ & $< -0.985                  $ & $< -0.967   					$ & $< -0.980                  $ & $< -0.944                  $\\
			
			{\boldmath$\Omega_{\phi 0}  $} & $0.701^{+0.013}_{-0.013} $  &  $0.702^{+0.013}_{-0.013}   $ & $0.701^{+0.012}_{-0.013}   $ & $0.701^{+0.013}_{-0.013} 		$ & $0.701^{+0.013}_{-0.013}   $ & $0.701^{+0.013}_{-0.013}   $\\
			
			{\boldmath$h              $} & $0.687^{+0.010}_{-0.010} $  &  $0.687^{+0.010}_{-0.010}   $ & $0.687^{+0.010}_{-0.010}   $ & $0.685^{+0.011}_{-0.011} 		$ & $0.686^{+0.011}_{-0.011}   $ & $0.686^{+0.010}_{-0.010}   $\\
			
			{\boldmath$V_0/\rho_c     $} & $0.354^{+0.015}_{-0.011} $  &  $0.705^{+0.016}_{-0.015}   $ & $< 0.0380				    $ & $0.660^{+0.046}_{-0.056}  		$ & $0.644^{+0.060}_{-0.071}   $ & $<5280				      $\\
			
			{\boldmath$\phi_i       $} & $< 0.290                   $ &  $-							 $ & $<156			            $ & $0.0030^{+0.0045}_{-0.0026}		$ & $< 0.000235                $ & $-						  $\\
			
			{\boldmath$\log_{10}{\phi'_i}$} & $-6.11^{+0.60}_{-0.53}$ &  $-6.02^{+0.64}_{-0.62}     $ & $-6.04^{+0.67}_{-0.62}      $ & $-4.12^{+0.56}_{-0.62}    		$ & $-3.98^{+0.62}_{-0.67}     $ & $-1.72^{+0.70}_{-0.66}     $\\
			
			{\boldmath$f              $} & $> 0.237$                  &   $-						$ & $-						    $ & $-								$ & $-						   $ & $-						  $\\
			
			{\boldmath$n              $} & $-$                  	  &   $-						$ & unconst.				 & $ < 0.0983 						$ & $< 0.163                   $ & $-						  $\\
			
			{\boldmath$\lambda_1        $} &  $-					$     &   $-                        $  & $-						    $ & $-								$ & $-					$ & $<614		               $\\
			
			{\boldmath$\lambda        $} &   $-					$     & $< 0.302                   $  & $-						    $ & $-								$ & $-						   $ & $< 0.465                   $\\
			
			{\boldmath$\log_{10}{\mu}            $} &   $-						$ & $-						    $ & $-							$ & $-						   		$ & $-						   $ & $-2.6^{+1.3}_{-1.4}        $\\
			
			\hline
			{$\Delta\textrm{AIC}_c$} &		$4.71$	& $2.32$ & $4.71$ & $2.35$ & $2.32$ & $4.73$ \\
			\hline
			{$\Delta$DIC} &		$0.58$	& $0.62$ & $0.53$ & $3.30$ & $3.41$ & $1.89$ \\
			\hline
		\end{tabular}
	\end{table*}
}
{\renewcommand{\arraystretch}{2.}%
	\begin{table*}
		\caption{Same as \Cref{tab:results}, obtained using the low-z+P16 datasets.}
		\label{tab:results_noH0}
		\begin{tabular} { l  c c c c c c }
			\hline
			&  PNGB  & EXP  & PL & IPL & SUGRA & DEXP \\ 
			\hline
			{\boldmath$w_0            $} & $< -0.978                 $  &  $< -0.985                  $ & $< -0.983                  $ & $< -0.951    					$ & $< -0.970                  $ & $< -0.956                   $\\
			
			{\boldmath$\Omega_{\phi 0}  $} & $0.695^{+0.014}_{-0.014} $  &  $0.695^{+0.014}_{-0.014}   $ & $0.695^{+0.014}_{-0.014}   $ & $0.693^{+0.014}_{-0.014}  		$ & $0.694^{+0.014}_{-0.014}   $ & $0.694^{+0.014}_{-0.015}   $\\
			
			{\boldmath$h              $} & $0.682^{+0.011}_{-0.011} $  &  $0.682^{+0.011}_{-0.011}   $ & $0.682^{+0.011}_{-0.011}   $ & $.679^{+0.012}_{-0.013} 		$ & $0.679^{+0.012}_{-0.012}   $ & $0.680^{+0.011}_{-0.011}   $\\
			
			{\boldmath$V_0/\rho_c     $} & $0.352^{+0.018}_{-0.013}$  &  $0.699^{+0.017}_{-0.017}   $ & $< 0.042				    $ & $0.638^{+0.060}_{-0.066}  		$ & $0.620^{+0.074}_{-0.080}   $ & $<5580					  $\\
			
			{\boldmath$\phi_i       $} & $< 0.334                   $ &  $-							 $ & $< 149            $ & $0.0025^{+0.0032}_{-0.0023}		$ & $< 0.000367                $ & $-						  $\\
			
			{\boldmath$\log_{10}{\phi'_i}$} & $-6.10^{+0.62}_{-0.55}     $ &  $-6.03^{+0.65}_{-0.62}      $ & $-6.03^{+0.67}_{-0.64}     $ & $-3.97^{+0.58}_{-0.64}    		$ & $-3.84^{+0.66}_{-0.71}     $ & $-1.76^{+0.74}_{-0.66}     $\\
			
			{\boldmath$f              $} & $> 0.233$                  &   $-						$ & $-						    $ & $-								$ & $-						   $ & $-						  $\\
			
			{\boldmath$n              $} & $-$                  	  &   $-						$ & unconst.				 & $ < 0.152 						$ & $< 0.243	                   $ & $-						  $\\
			
			{\boldmath$\lambda_1        $} &  $-					$     &   $-                        $  & $-						    $ & $-								$ & $-						   $ & $<634				  $\\
			
			{\boldmath$\lambda        $} &   $-					$     & $< 0.314                    $  & $-						    $ & $-								$ & $-						   $ & $<0.536					  $\\
			
			{\boldmath$\log_{10}{\mu}            $} &   $-						$ & $-						    $ & $-							$ & $-						   		$ & $-			   $ & $-2.5^{+1.3}_{-1.5}        $\\
			
			\hline
			{$\Delta\textrm{AIC}_c$} &		$4.71$	& $2.31$ & $4.71$ & $2.36$ & $2.33$ & $4.72$ \\
			\hline
			{$\Delta$DIC} &		$0.20$	& $0.13$ & $0.13$ & $2.70$ & $2.67$ & $1.14$ \\
			\hline
		\end{tabular}
	\end{table*}
}
{\renewcommand{\arraystretch}{2.}%
	\begin{table*}
		\caption{Same as \Cref{tab:results}, obtained using the low-z+R18 datasets. While is not shown explicitly, here $r_d$ is fitted as a free parameter, for which we find very consistent values of $r_d \sim 142 \pm 5$ [Mpc] at $2\sigma$, for all models.} 
		\label{tab:results_low}
		\begin{tabular} { l  c c c c c c }
			\hline
			&  PNGB  & EXP  & PL & IPL & SUGRA & DEXP \\ 
			\hline
			{\boldmath$w_0            $} & $< -0.906                  $  &  $< -0.918               $ & $< -0.919                 $ & $-0.865    					$ & $< -0.914                   $ & $< -0.866                   $\\
			
			{\boldmath$\Omega_{\phi 0}  $} & $0.718^{+0.025}_{-0.026} $  &  $0.718^{+0.025}_{-0.025}  $ & $0.718^{+0.025}_{-0.026}   $ & $0.729^{+0.032}_{-0.030}  		$ & $0.739^{+0.036}_{-0.034}   $ & $0.723^{+0.027}_{-0.028}   $\\
			
			{\boldmath$h              $} & $0.715^{+0.024}_{-0.024} $  &  $0.715^{+0.024}_{-0.023}  $ & $0.716^{+0.024}_{-0.024}   $ & $0.713^{+0.024}_{-0.024} 		$ & $0.714^{+0.023}_{-0.024}   $ & $0.714^{+0.024}_{-0.024}   $\\
			
			{\boldmath$V_0/\rho_c     $} & $0.379^{+0.091}_{-0.044}$  &  $0.736^{+0.096}_{-0.058}   $ & $< 0.111 				   $ & $0.626^{+0.077}_{-0.057}  		$ & $0.82^{+1.10}_{-0.71}       $ & $< 53100                  $\\
			
			{\boldmath$\phi_i       $} & $< 0.667                   $ &  $-							$ & $< 145            $ & $0.0021^{+0.0016}_{-0.0011}		$ & $< 0.0391                $ & $-						  $\\
			
			{\boldmath$\log_{10}{\phi'_i}$} & $-5.86^{+0.79}_{-0.74} $ &  $-5.77^{+0.88}_{-0.86}    $ & $-5.79^{+0.85}_{-0.84}     $ & $-3.45^{+0.81}_{-0.88} 		$ & $-2.70^{+1.30}_{-1.30}      $ & $-1.40^{+1.00}_{-1.10}        $\\
			
			{\boldmath$f              $} & $> 0.243$                  &   $-						$ & $-						    $ & $-								$ & $-						   $ & $-						  $\\
			
			{\boldmath$n              $} & $-$                  	  &   $-						$ & unconst.				 & $ < 0.508 						$ & $< 2.35	                   $ & $-						  $\\
			
			{\boldmath$\lambda_1        $} &   $-					$     & $-                      $  & $-						    $ & $-								$ & $-						   $ & $< 942                    $\\
			
			{\boldmath$\lambda        $} &   $-					$     & $< 0.709                    $  & $-						    $ & $-								$ & $-						   $ & $<0.871      $\\
			
			{\boldmath$\log_{10}{\mu}            $} &   $-						$ & $-						    $ & $-							$ & $-						   		$ & $-						   $ & $-3.1^{+2.2}_{-1.9}        $\\
			
			\hline
			{$\Delta\textrm{AIC}_c$} &		$4.70$	& $2.25$ & $4.67$ & $2.01$ & $1.58$ & $4.56$ \\
			\hline			
			{$\Delta$DIC} &		$0.27$	& $0.08$ & $0.05$ & $1.06$ & $1.98$ & $0.95$ \\
			\hline
		\end{tabular}
	\end{table*}
}

\end{document}